\documentclass[11pt]{article}
\usepackage{epsfig}
\newcommand{\ct}[1]{$^{\cite{#1}}$}

\newcommand{\bee}{\begin{equation}}
\newcommand{\ee}{\end{equation}}
\newcommand{\beea}{\begin{eqnarray}}
\newcommand{\eea}{\end{eqnarray}}

\begin{document}
\title{An Interesting Fitting Formula of Quark Masses}
\author{{Da Qing Liu, Ji Ming Wu}\\
        {\small Institute of High Energy Physics, Chinese Academy
                of Sciences, P.R. China}\\
       }
\maketitle
\begin{center}
\begin{minipage}{5.5in}
%\vskip 0.3in
{\bf Abstract}\\
We show an interesting empirical formula of quark masses here, which is
found by implementing a least squares fit. In this formula
the measured QCD coupling is almost a "best fitting coupling".
\end{minipage}
\end{center}
%\vskip 0.5in

\noindent

There are many papers working on the masses of quarks, in theory
or in experiment.  For theory, the difficulty is, masses of quarks
can not been determined by first principle. The masses of known
six quarks correspond to six free parameters in standard model
(SM). On the other hand, there appears a mystery problem in the
SM, that is, comparing to \(t\) quarks, the other five quarks can
be considered as ones with vanishing mass. However, there are
empirical formulae to illuminate this hierarchy problem, such as
reference \cite{0406}. We show another interesting mass
fitting for the known six quarks in this note, which can be
compared with the results in \cite{0406}. Although it is  empirical,
this formula is significant agreement with experiment
values. We expect this formula can help us to discover the
correct theory of flavor. A byproduct of this formula is,
that the best fitting coupling is just the experiment one.

We first list the \(\overline{MS}\) quark masses in table 1. These
data are from reference \cite{pdg} in different scale. Because
pole masses of light quarks are physical meaningless and pole masses of heavy
quarks also have evil definitions\ct{9804275}, we do not use the data of pole
mass here.

\begin{table}[ht]
\begin{center}
\begin{tabular}{|c|c|c|c|}
  \hline
          & $n_g=1$ & $n_g=2$ & $n_g=3$ \\ [0.5ex] \hline
  $n_s=1$ & $3.25\pm 1.75Mev(2Gev)$ & $1.2\pm 0.2Gev({m_c})$
  & $174\pm 5 Gev^\dag$ \\ \hline
  $n_s=2$ & $7.0\pm 2.0Mev(2Gev)$ & $115\pm 35 Mev(2Gev)$ &
  $4.25\pm 0.2Gev({m_b})$ \\ \hline
\end{tabular} \\  [0.5ex]
\begin{minipage}{5.5in}
\emph{Table 1:} Masses of quarks in $\overline{MS}$ scheme from
\cite{pdg}. $n_g$ is the number of generator and $n_s$ is the
number of "isospin": $n_s=3(1-|Q|)$, where $Q$ is the charge of
quark. Since reference \cite{pdg} only gives mass ranges of
different quarks, we take its midpoint as our input. For
instance, the mass of $u$ quark $3.25\pm 1.75Mev(2Gev)$ just
corresponds to the range $1.5Mev\leq m_u(2Gev)\leq 5 Mev$ in
reference \cite{pdg}. The quantities in
bracket are scales where we obtained quark masses. \\
$^\dag$: This is the pole mass of top quark, which should be
converted into $\overline{MS}$ mass using Eqn. (\ref{convt})\ct{pdg}.
\end{minipage}
\end{center}
\end{table}

We then renormalize different quark masses to the mass of $Z$ particle
$m_Z=91.2Gev$ using renormalization group equation. Relevant
formulae have been shown in reference \cite{fusaoka}. We take here
$\alpha_s(m_Z)=0.117$\ct{pdg}, which corresponds
$\Lambda^{(5)}=197Mev,\,\Lambda^{(4)}=274Mev$ and
$\Lambda^{(3)}=310Mev$. The results have been shown in Table 2. As
argued in Tab. 1, one should convert the pole mass of \(t\) quark
into $\overline{MS}$ mass. Since $m_t>>m_q$, where
$q=u,\,d,\,s,\,c,\,b$, the mass is converted by a simpler
relation, \bee \label{convt}
m^{pole}=m(m)[1+\frac{4\alpha_s(m)}{3\pi}+(13.44-1.04\times 5)
({\alpha_s(m)\over \pi})^2]. \ee

\begin{table}[ht]
\begin{center}
\begin{tabular}{|c|c|c|c|}
  \hline
          & $n_g=1$ & $n_g=2$ & $n_g=3$ \\ [0.5ex] \hline
  $n_s=1$ & 0.00019(10) & 0.597(100) & 173(5) \\ \hline
  $n_s=2$ & 0.00041(12) & 0.067(21) & 2.94(17) \\
 \hline
\end{tabular} \\  [0.5ex]
\emph{Table 2:} The $\overline{MS}$ masses ($Gev$) at scale $m_Z$.
\end{center}
\end{table}

In QCD, the contributions of quark mass to high energy parameters,
such as anomalous dimension functions, decay behaviors beyond tree
level, coupling, and even more, the evolutions of masses
themselves, have the form $\ln{{m\over m_0}}$, where $m_0$ is some
subtraction mass. Thus, we consider here the relations of the
logarithm of quark mass, for instance, $\ln{m/m_0}$, where we
choose $m_0=1Gev$. The results are listed in Table 3.

\begin{table}[ht]
\begin{center}
\begin{tabular}{|c|c|c|c|}
  \hline
          & $n_g=1$ & $n_g=2$ & $n_g=3$ \\ [0.5ex] \hline
  $n_s=1$ & -6.26(54) & -0.515(167) & 5.155(29) \\ \hline
  $n_s=2$ & -5.49(29) & -2.696(304) & 1.079(59) \\
 \hline
\end{tabular} \\  [0.5ex]
\emph{Table 3:} $y(n_g,n_s)=\ln{m/m_0}$ at scale $m_Z$, where
$m_0=1Gev$.
\end{center}
\end{table}

We find that, for definite $n_s$, there is an approximate linear
relation among different $n_g$. One can use formula $a_1 n_s
n_g+a_2 n_g+a_3 n_s+a_4$ to fit quark masses. We use the least
squares method with wight to obtain $a_i$. That is, find
parameters $a_i$ to minimize function \bee
f=\sum\limits_{n_g=1,2,3, \\ n_s=1,2} (y(n_g,n_s)-(a_1 n_s n_g+a_2
n_g+a_3 n_s+a_4))^2 w(n_s,n_g), \ee where $w(n_s,n_g)$ is the
wight: $w(n_s,n_g)=dy^{-2}(n_s,n_g)$, where $dy$ is the error of
$y=\ln{m/m_0}$. The coefficients are
$a_1=-2.31(20),\,a_2=7.99(32),\,a_3=2.85(58),$ and
$a_4=-14.73(94)$.

Notice that \bee a_1 n_s n_g+a_2 n_s+a_3 n_g+a_4=(a_1
n_g+a_3)(n_s+{a_2\over a_1})+(a_4-{a_2a_3\over a_1}),
\label{con}\ee a more convenient approach is to redefine $m_0$ and
then fit mass using $(c_1 n_g+c_2)(n_s+c_3)$. From Eqn. \ref{con}
we let $m_0$ satisfy $\ln{\frac{m_0}{1Gev}}=a_4-{a_2a_3\over
a_1}=-4.886$, or $m_0(m_Z)=7.55Mev$. Data of $\ln{m/m_0}$ have
been shown in Table 4.

\begin{table}[ht]
\begin{center}
\begin{tabular}{|c|c|c|c|}
  \hline
          & $n_g=1$ & $n_g=2$ & $n_g=3$ \\ [0.5ex] \hline
  $n_s=1$ & -1.375(538) & 4.372(167) & 10.041(30) \\ \hline
  $n_s=2$ & -0.608(286) & 2.191(304) & 5.965(59) \\
 \hline
\end{tabular} \\  [0.5ex]
\emph{Table 4:}  $y(n_g,n_s)=\ln{m/m_0}$ at scale $m_Z$, where
$m_0=0.00755Gev$.
\end{center}
\end{table}

Using the least squares method, that is, minimizing the function
\bee \label{fit1} \sum\limits_{n_g=1,2,3, \\ n_s=1,2}
(y(n_g,n_s)-(c_1 n_g+c_2)(n_s+c_3))^2 w(n_s,n_g), \ee one obtains
$c_1=-2.311(25),\,c_2=2.848(28)$ and $c_3=-3.459(35)$. In fact,
one can use a more symmetrical form,
$c_1(n_g+c_2^\prime)(n_s+c_3)$, where $c_2^\prime=c_2/c_1$, to fit
$y(n_g,n_s)$. Notice $c_3\simeq 3c_2/c_1$, one can furthermore
reduce parameters $c1,\,c2,\,c3$ into two parameters $c1,\,c2$, if
he performs constraint $c_3=3c_2/c_1=3c_2^\prime$. We did not do
it here. Quark masses and the fitting formula are plotted in Fig.
1. This empirical formula can be compared with the results
introduced in ref. \cite{0406}.

To discuss the quality of fitting in Eqn. (\ref{fit1}), we study
modified coefficient of determination for statistics:
 \bee R^2=1-\frac{SSE/N_E}{SST/N_T}. \label{mr}\ee
In Eqn. (\ref{mr}) $N_E=6-3-1=2$ is the degree of sum squared
error (SSE),
$SSE=\sum\limits_{n_s,n_g}(y(n_g,n_s)-\hat{y}(n_g,n_s))^2$, where
\(\hat{y}\) is the fitting value of $y$, $N_T=6-1-1=4$ is the
degree of total sum of squares (SST),
$SST=\sum\limits_{n_s,n_g}(y(n_g,n_s)-\bar{y})^2$, where
\(\bar{y}\) is the average of $y$: \(\bar{y}={1\over
6}\sum\limits_{n_s,n_g}y(n_g,n_s)\). The additional subtraction of
1 is due to subtraction constance $m_0$. In other literature $R^2$
is written as $\bar{R}^2$.

Generally, $0\leq R^2\leq 1$. One obtains here that $R^2=0.99584$,
which is very close to \(1\). This roughly means that, about 99.6
percent of the mass statistics can be interpreted by this
empirical fitting. Or, the part which can not been interpreted by
the fitting is no more than 1\%. We conclude that this fitting is
a quite good empirical formula.

All the calculations given above depend on the coupling
\(\alpha_s(m_Z)\). But, \(\alpha_s(m_Z)\) itself is determined by
experiment and it also has error. Therefore, it is interesting to
study the fitting behavior at different \(\alpha_s(m_Z)\). Due to
the experiment interesting, we vary \(\alpha_s(m_Z)\) from 0.09 to
0.13 here. We repeat all the calculations, taking Tab. 1 as input,
and then study the behavior of modified coefficient of
determination $R^2$ for the fitting formula (\ref{fit1}),
(\ref{mr})\footnote{Here, for simplification, we just make an
assumption that the data in Tab. 1 is irrelvant with
\(\alpha_s(m_Z)\).}. The results are shown in Fig 2. One just
finds that \(R^2\) approaches its maximum at the range $0.115<
\alpha_s(m_Z)<0.12$.  We call this coupling as best fitting
coupling $\alpha_s^b$, where the fitting formula (\ref{fit1})
works best. We use $d\alpha_s^b=\sqrt
{\sum\limits_{n_s,n_g}({\partial \alpha_s^b /
\partial m(n_s,n_g) })^2 dm^2(n_s,n_g)}$, where $m(n_s,n_g)$ is the data in Tab.
1 and $\partial \alpha_s^b / \partial m(n_s,n_g)$ can be extracted
by a small shifting of \(m(n_s,n_g)\), to estimate error of
\(\alpha_s^b\), \(d\alpha_s^b\). The result is
$\alpha_s^b=0.118(31)$. The big error is mainly due to the errors
of $u,\,d,\,s$ quarks masses.

We see that the measured coupling is very close to the best
fitting coupling. Here is a possible interpretation. As we know,
the masses in Tab. 1 are mean ones, which are extracted by
connecting various theories and experiments. In this sense, we say
that the data in Tab. 1 is unprejudiced estimation of the true
masses of quarks, as long as we have performed enough estimations.
Suppose QCD and renormalization theory are both right theories and
remember that the mean measured coupling \(\alpha_s(m_Z)=0.117\)
is just the estimation of the true coupling \(\alpha_s^t\). We
conclude that, if fitting equation (\ref{fit1}) is a right or an
approximate right behavior, \(\alpha_s^b\) should also be a
estimation to \(\alpha_s^t\), although \(\alpha_s^t\) itself
depends on the level of loops calculations. On one hand, if the
behavior of $y(n_g,n_s)$ is complete random or is not linear at
all, or in other words, the fitting (\ref{fit1}) is not a correct
one, one should obtain two bad results, one is that $R^2$ is not
so close to 1, the other is that, generally, \(\alpha_s^b\neq
\alpha_s^t\), unless by chance. Since $R^2$ is very close to
unitary, we expect the equation (\ref{fit1}) is a right or an
approximate right behavior of quarks. On the other hand, when one
says that the fitting equation (\ref{fit1}) is a right or an
approximate right behavior, he always implies that this statement
is obtained at correct coupling, \(\alpha_s^t\). This means that,
when the coupling deviates away \(\alpha_s^t\), the fitting will
go to bad. Or, in other words, the \(\alpha_s^t\) should be equal
to the best coupling, \(\alpha_s^b\), provided one takes correct
mass input in Tab. 1. Therefore, \(\alpha_s^b\) is also an
estimation of \(\alpha_s^t\). It is understood that
\(\alpha_s^b\simeq \alpha_s\).

From Fig. 1 and $R^2$ check, the linear fitting is quite good
agreement with the experiment data. \(s\) quark
lies a little below the fitting line, (or on the contrary, \(d\)
quark lies a little above the fitting line,) which may be
considered as the correction due to QED and statistic error. In
fact, if one uses linear fitting to fit the masses of leptons,
which have $|Q|=1$, the experiment of \(\mu\) lepton should lies
above the fitting line( Since the lepton does not enjoy strong
interaction, we do not discuss the fitting for lepton masses in
detail here).

If the logarithm of quark masses does have linear behavior, there
are some interesting infers immediately.

For instance, one can use this formula to extract masses of four
generator quarks. For the heavier quark $t^\prime$, we get
$m_{t^\prime}(m_Z)=51(17)Tev$, which is beyond our experiment
capability. But the searching of lighter quark $b^\prime$ is not
beyond our experiment capability. In fact, using extraction of the
linear fitting, we obtain $m_{b^\prime}(m_Z)=85(23)Gev$, which
corresponds to the pole mass $m^{pole}_{b^\prime}=91(25)Gev$. In
other words, $m^{pole}_{b^\prime}\approx m_Z$ dramatically.

It is also a puzzle that whether the fourth-generate quarks exist
or not. Until nowadays, we did not find the fourth-generate
quarks. But since $m^{pole}_{b^\prime}\approx m_Z$, one should
check the data at the vicinity around $m_Z$ more carefully.
According to reference \cite{olivar}, the mass gap between the
fourth-generate quarks is so large that possibly there is no
fourth-generate quark at all.

It is hard to understand why top quark has so large mass in SM. In
some seesaw mechanism, for instance, the source of top quark is
significant different with that of other quarks\ct{9703249} and
therefore the formula of top quark is also different with that of
other quarks. But we see here that the mass of top quark extracted
from the linear behavior of the logarithm of quark mass agrees
well with the measured ones. This implies that the top quark is
also a "common" quark and the masses of top quark and all the other known quarks
share the same source, although $m_t>>m_Z$. We expect this is helpful to
understand the source of particle masses and correct theory of
flavor.

In some references, the vanishing of up quark mass is used to
solve strong CP puzzle in QCD. However, if the fitting equation
(\ref{fit1}) is right or approximate right, the mass of up quark
is never vanishing. This means that the solution of CP broken in
QCD is not the vanishing of up quark mass. It is possibly from
other mechanism, for instance, U(1) symmetry.

At last, the subtract
mass is $m_0(m_Z)=7.55Mev$, which corresponds $m_0(m_0)=428Mev$,
or roughly equals to constituent light quarks mass.

In summary, the significant agreement of equation (\ref{fit1})
shows that the masses of known quarks are never random. Therefore,
equation (\ref{fit1}) should be included in the full theory. We
expect this equation should give some clue of full theory, such as
the source of masses or flavor physics.

The author D.Q. Liu is very grateful to Prof. J. Ferrandis for reading
the manuscript and helpful comments.

%\newpage

\begin{figure}[ht]
\centerline{\epsfig{file=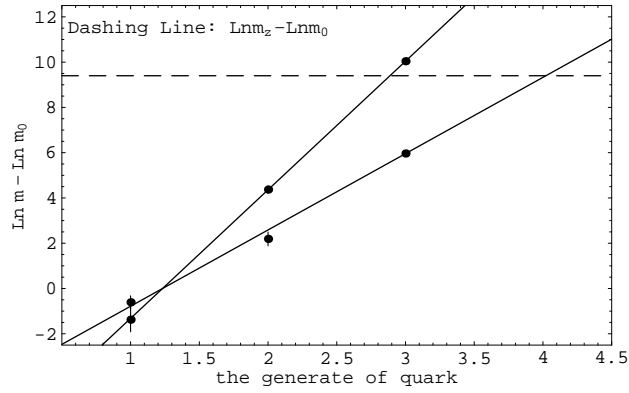,width=10.0cm,clip=}}
\caption[]{$\ln(\frac{m_{n_g}}{m_0})$ vs. $n_g$, where $n_g$
stands for the generator of quarks while $m_0$ is
pointed out in Tab. 4. The dashing line is $\ln({m_Z \over m_0})$.}
\label{fig:mass1}
\end{figure}

\begin{figure}[ht]
\centerline{\epsfig{file=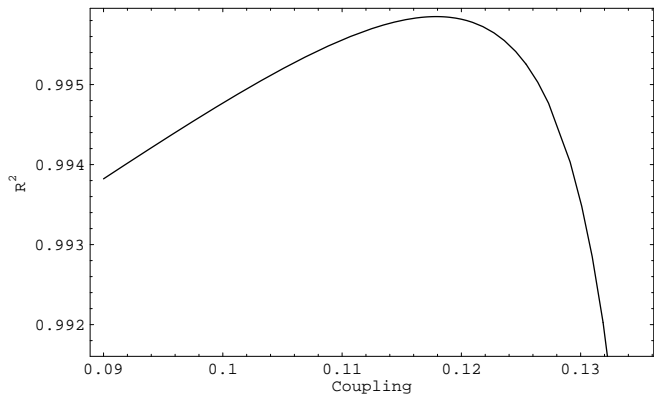,width=10.0cm,clip=}}
\caption[]{$R^2$ vs. $\alpha_s(m_Z)$.}
\label{fig:mass2}
\end{figure}

\end{document}